\newcommand{\be}{\begin{equation}}
\newcommand{\ee}{\end{equation}}
\newcommand{\ba}{\begin{eqnarray}}
\newcommand{\ea}{\end{eqnarray}}

\newcommand{\Tr}{\hbox{Tr}}
\documentclass[doublespacing]{elsart}
\usepackage{amssymb}
\usepackage{amsmath}
\begin{document}
\begin{frontmatter}
\title{Dynamical origin of the \\$\star_{\theta}$- noncommutativity in field theory  from quantum mechanics}
\author{Marcos Rosenbaum, J. David Vergara and L. Rom\'an Ju\'arez }
\address{Instituto de Ciencias Nucleares, UNAM, A. Postal 70-543,
M\'exico D.F.}
\begin{abstract}
We show that introducing an extended Heisenberg algebra in the context of the
Weyl-Wigner-Groenewold-Moyal formalism leads to a deformed
product of the classical dynamical variables that is inherited to the level of
quantum field theory, and that allows us to relate the operator space
noncommutativity in quantum mechanics to the quantum group inspired
algebra deformation noncommutativity in field theory.
\end{abstract}

\begin{keyword}
Noncommutativity \sep star-products

\PACS 02.40Gh \sep 11.10.Nx
\end{keyword}
\end{frontmatter}

\section{Introduction}

Theoretical physics has provided us a fairly deep understanding of
the microscopic structure of matter,
but very little is known regarding the microscopic structure of space-time.\\
From a methodological point of view, the use of a noncommutative
structure for space-time coordinates had already been proposed in
the early days of field theory as a failed hope at finding an
effective and Lorentz invariant cutoff needed to control the
ultraviolet divergences plaguing the theory. From a conceptual and
theoretical point of view there is a simple heuristic argument -
based on Heisenberg's Uncertainty Principle, the Einstein
Equivalence Principle and the Schwarzschild metric - which shows
that the Planck length seems to be a lower limit to the possible
precision measurement of position, and that shorter distances do not
appear to have an operational meaning \cite{doplicher}. Thus Quantum
Mechanics and Field Theory, at dimensions of the order of the Planck
length, ought to incorporate in their very structure the
noncommutativity of space-time by replacing the concept of a
space-time point by a cell of a dimension given by the Planck scale
area. Under these premises the very concept of manifold as an
underlying mathematical structure of physical theories becomes
questionable and some people are convinced that a new paradigm of
geometrical space is needed. The noncommutative geometry of Connes
\cite{connes}, which by resorting to arbitrary and noncommutative
$C^*$-algebras dualizes geometry and replaces its usual notions of
manifolds and points by a new calculus based on operators in Hilbert
space and the use of spectral analysis, epitomizes this line of
thought. More recently there has been further evidence of space-time
noncommutatitvity \cite{seiberg} coming from certain models of
string theory which, although with a geometry quite different from
that of noncommutative geometry is not incompatible with it, and has
led to the same issue of noncommutativity of space-time at short distances. \\
In the noncommutative quantum field theory rooted on the
phenomenology of the low energy approximation of string theory in
the presence of a strong magnetic background, the fields on a target
space of space-time canonical coordinates are replaced by a
$C^*$-algebra of functions with a deformed product given by the so
called Groenewold-Moyal star-product: \be f(x)\star_{\theta} g(x)=
f(x) e^{(\frac{i}{2}{\overleftarrow \partial_{i}}
\theta^{ij}{\overrightarrow\partial_{j}})} g(x),\label{1.1} \ee
where the constant real and invertible anti-symmetric tensor
$\theta^{ij}$ has dimensions of length squared. One interpretation (see {\it
e.g.} \cite{alvarez}) for the origin of this noncommutativity is
based on postulating the replacement of the space-time argument of
canonical coordinates $x^i$  of field operators by a ``space-time"
of Hermitian operators obeying the Heisenberg algebra \be [{\hat
x}^{i} , {\hat x}^{j}]=iI\theta^{ij},\;\;\;
i,j=1,\dots,2d\label{1.2} \ee where $I$ is an identity operator.
Operators ${\mathcal O}({\hat x})$, acting on a Hilbert space of
delta-function normalizable functions in d-dimensions, are then
defined in terms of the basic operators (\ref{1.2}) by means of the
Weyl basis $g(\alpha,{\hat x})=e^{i\alpha_i {\hat x}^i}$.  Using now
the Weyl-Moyal correspondence \be {\mathcal O}({\hat x}) = \int
d^{2d} \alpha \;g(\alpha, {\hat x}) {\tilde O}_W(\alpha),
\label{1.4} \ee where ${\tilde O}_W(\alpha)$ is the Fourier
transform of the Weyl function corresponding to ${\mathcal O}$, it
follows, in complete analogy to the results derived from the
Weyl-Wigner-Groenewold-Moyal (WWGM) formalism of quantum mechanics
(see the following section), that the Weyl function corresponding to
the operator product ${\mathcal O}_{1}{\mathcal O}_{2}$ is given by
\be (O_1)_W \;\star_{\theta} (O_2)_W.\label{1.5} \ee For a review of
noncommutative quantum field theory based on these criteria see,
{\it e.g.}, \cite{szabo}.

An alternative and Lorentz invariant (in the twisted symmetry sense)
interpretation of the origin of the star-product (\ref{1.1}) comes
from considering the twisted coproduct of the Hopf algebra
${\mathcal H}$ of the universal enveloping ${\mathcal U({\mathcal
P})}$ of the Poincar\'e algebra ${\mathcal P}$. It can be shown (see
{\it e.g.} \cite{chaichian}) that for a certain Drinfeld twisting of
the coproduct with an invertible ${\mathcal F}\in {\mathcal
U({\mathcal P})}\otimes {\mathcal U({\mathcal P})}$ such that \be
{\mathcal F}_{12}(\Delta\otimes id){\mathcal F}={\mathcal
F}_{23}(id\otimes\Delta),\ \ \  (\epsilon\otimes id){\mathcal
F}=1=(id\otimes\epsilon){\mathcal F},\label{1.6} \ee this coproduct
induces a deformation in the product, $m\rightarrow m_{\mathcal F}$,
of the module algebra ${\mathcal A}=C^\infty (M)$ over ${\mathcal
H}$, such that the action of ${\mathcal H}$ on ${\mathcal A}$
preserves covariance, {\it i.e.} \be h\triangleright m_{\mathcal
F}(a\otimes b)= m\circ [({\mathcal F}_{(1)}^{-1}\:\; \triangleright
a)\otimes ({\mathcal F}_{(2)}^{-1}\:\; \triangleright b)]=
a\star_{\theta}b,\label{1.7} \ee where $a,b \in {\mathcal A}$ and
$h\in {\mathcal H}$, and we have used the Sweedler notation
throughout. In particular, considering the coordinates $x^i$ as
elements of ${\mathcal A}$, equation (\ref{1.7}) implies that \be
[x^i , x^j]_{\star_\theta} \equiv x^i \star_\theta x^j - x^j
\star_\theta x^i=i\theta^{ij}.\label{1.8} \ee Note, however, that
although both of the above described representative lines of thought
lead to the same algebra of operators for noncommutative quantum
field theory, the origins of this noncommutativity appear to be
quite different. In the later case, as has been stressed by
Chaichian {\it et al.}, the product (\ref{1.8}) is inherited from
the twist of the operator product of quantum fields and no
noncommutativity of the coordinates was used in the derivation of
(\ref{1.7}); while in the line of thought described in
\cite{alvarez} the assumed noncommutativity of the space-time
operators forms an essential part of the ensuing arguments. However,
the inference that the multiplication in the algebra of fields is
given by the star-product (\ref{1.7}) is an external ingredient
imported from the phenomenology of string theory.

 Since quantum mechanics is strongly interwoven into noncommutative geometry,
and since single particle quantum mechanics can be seen, in the free
field or weak coupling limit, as a mini-superspace sector of quantum
field theory where most degrees of freedom have been frozen   ({\it
i.e,} as a one-particle sector of field theory), it is suggestive
that a further study of quantum mechanics  in this noncommutative
context, and in particular in the WWGM formalism based on a
Heisenberg algebra extended to incorporate space noncommutativity,
may help to shed some additional light on the origins of the product
(\ref{1.8}) in the algebra of noncommmutative field theory.

Observe, however, that in the strict sense of quantum mechanics only
expectation values have a physical meaning. This, in the WWGM
quantum formalism, translates to the fact that the $c$-equivalent of
a quantum operator, or to that effect of a product of operators,
appears together with the Wigner quasi-distribution function inside
of a phase-space integral. In the case of the standard Heisenberg
algebra of usual quantum mechanics, the Wigner function is the same
as the Weyl equivalent of the von Neumann density matrix and the
Weyl equivalent of a product of operators (given by the
Groenewold-Moyal product of their respective Weyl equivalents) is
indeed the $c$-function that would appear in the integrand
multiplying the Wigner function. On the other hand, as it is shown
in the next section, this is not true for the case of a quantum
mechanics with an extended Heisenberg algebra. In fact, as shown in
equations (27) or (28) there, either of which can be used to
evaluate the expectation value of a product of operators, the Weyl
equivalent of a product of operators (given by (30) with a composite
$\bigstar$-product defined by (25), (31) and (32) ) is not the one
required in the integrands in order to arrive at the correct
expectation values. Hence this $\bigstar$-product does not appear as
a natural ingredient of the quantum mechanical formalism when
considering only Schr\"{o}dinger operators.

The purpose of this work is to show nonetheless that when
considering in addition Weyl equivalents of Heisenberg operators,
the $\star_{\theta}$ product for the algebra of what can then be
identified as canonical dynamical variables, emerges naturally
within the theory and thus allows for a further link between the
points of view of quantum operator space noncommutativity, as
presented in \cite{alvarez}, and the quantum group inspired algebra
deformation noncommutativity, discussed in \cite{chaichian}. Lastly
we could expect as well that a detailed study of exactly solvable
models in the frame of this extended Heisenberg algebra WWGM
formalism may also be helpful to achieve a further understanding of
the possible phenomenological consequences in space of the
noncommutativity in field theory. In this context, the above
observations as well as some additional ones contained below are
also pertinent to some works that have appeared recently in the
literature on what has been called noncommutative quantum mechanics.

\section{Quantum Mechanics on Extended Heisenberg Algebras in the WWGM Formalism}

By an extended Heisenberg algebra we understand the algebra of
position and momentum operators satisfying the commutation relations
\begin{align}
[{\hat R}_i, {\hat R}_j]  &= i\theta_{ij}, \label{noncomm1} \\
[{\hat P}_i, {\hat P}_j] &= i\hbar {\bar \theta}_{ij},\label{noncomm2} \\
[{\hat R}_i,{\hat P}_j] &= i \hbar \delta_{ij}, \label{noncomm3}
\end{align}
where ${\hat R}_i$,${\hat P}_i$\;\; $i=1,\dots,d$  are the
components of the position and momentum quantum operators,
respectively, with component eigenvalues on ${\mathbb R}^{d}$, and
$\theta_{ij}$ and ${\bar \theta}_{ij}$ are evidently antisymmetric
matrices, which in the most general case can be functions of the
generators of the above algebra. For our present purposes and
algebraic simplicity, in what follows we shall set ${\bar
\theta}_{ij}=0$ and $d=2$, and consider only the zeroth order
constant term of the Taylor expansion of $\theta_{12}\equiv \theta$.
(For $\theta$ constant, the formalism described below can be
generalized to include more spatial dimensions  in a fairly
straightforward way, and it also can be extended to incorporate
space-time noncommutativity  by parameterizing the time and
considering it as an extra variable. See for example \cite{Pinzul}).
From an intrinsically noncommutative operator point of view, the
development of a formulation for the quantum mechanics based on the
above extended Heisenberg algebra of operators requires first a
specification of a representation for the generators of the algebra,
second a specification of the Hamiltonian which governs the time
evolution of the system and last a specification of the Hilbert
space on which these operators and the other observables of the
theory act. As for the choice of the Hilbert space, a reasonable
assumption is that it can be taken to be the same as that for the
corresponding system in the usual quantum mechanics, but for a
realization of the extended Heisenberg algebra, because of the
noncommutativity (\ref{noncomm1}), we can not use configuration
space as a basis. We can use, however, for a basis either of the
eigenkets $|p_1,p_2\rangle$, $|q_1,p_2 \rangle$, $|q_2,p_1 \rangle$,
of the commuting pairs of observables $({\hat P}_1,{\hat P}_2)$,
$({\hat R}_1,{\hat P}_2)$, or $({\hat R}_2,{\hat P}_1)$,
respectively, or any combination of the
$(R,P)$ such that they form a complete set of commuting observables. \\
Having in mind generalizations to include the noncommutativity
(\ref{noncomm2}), we choose as the realization of our extended
Heisenberg algebra the one based on $|q_1,p_2 \rangle$. The
construction follows standard procedures ({\it cf.}\cite{messiah}):
Consider the unitary operator ${\hat S}(\gamma)=e^{î\gamma {\hat
R}_2}$ ($\gamma$ is an arbitrary parameter) and evaluate its
commutators with ${\hat R}_1$ and ${\hat P}_2$. It is easy to show
that \be {\hat S}(\gamma)|q_1,p_2 \rangle = |q_1 -\theta\gamma, p_2
+\hbar\gamma\rangle.\label{2.1} \ee Assuming now that $\gamma$ is an
infinitesimal and evaluating $\langle q_1,p_2| {\hat
S}(\gamma)|q^{\prime}_1,p^{\prime}_2 \rangle$
to first order in $\gamma$ results in\\
$$\langle q_1,p_2| {\hat R}_2 |q^{\prime}_1, p^{\prime}_2 \rangle
=(-i\theta\partial_{q_1} + i\hbar\partial_{p_2})\langle
q_1,p_2|q^{\prime}_1, p^{\prime}_2 \rangle,$$ so the realization of
${\hat R}_2$ in this basis is \be {\hat R}_2 =-i\theta\partial_{q_1}
+ i\hbar\partial_{p_2}.\label{2.2} \ee

Considering next  the unitary operator ${\hat
S}(\lambda)=e^{î\lambda {\hat P}_1}$ and following a similar
procedure we get \be {\hat P}_1 = -i\hbar\partial_{q_1}.\label{2.3}
\ee The representations for the remainder of the generators ${\hat
R}_1$ and ${\hat P}_2$ of the algebra are obviously simply
multiplicative. (Note that by making use of (\ref{2.1}) we can
readily make the change of basis
$|q_1,p_2\rangle\rightarrow|p_1,p_2\rangle$ and derive the
representations  ${\hat R}_1 = i\hbar\partial_{p_1}$ and ${\hat R}_2
= i\hbar\partial_{p_2} +\frac{\theta}{\hbar}p_1$ for the extended
Heisenberg algebra generators in the momentum representation. In
this case ${\hat P}_1$ and ${\hat P}_2$ are obviously just
multiplicative. All our calculations could then be related to that
basis.)
\\
For later calculations we shall be needing to evaluate the
transition function $\langle q_1,p_2|q_2,p_1\rangle$. This can be
derived \cite{acat} by noting that \be \langle q_1,p_2|{\hat R}_2
|q_2,p_1\rangle = q_2 \langle q_1,p_2|q_2,p_1\rangle
=i(\hbar\partial_{p_2} -\theta\partial_{q_1}\langle
q_1,p_2|q_2,p_1\rangle,\label{2.4} \ee and \be \langle q_1,p_2|{\hat
P}_1 |q_2,p_1\rangle = p_1 \langle q_1,p_2|q_2,p_1\rangle
=-i\hbar\partial_{q_1}\langle q_1,p_2|q_2,p_1\rangle.\label{2.5} \ee
Combining these two expressions yields \be (\hbar q_2 - \theta
p_1)\langle q_1,p_2|q_2,p_1\rangle=i\hbar\partial_{p_2} \langle
q_1,p_2|q_2,p_1\rangle,\label{2.6} \ee which can be readily solved
to give, after normalization, \be \langle
q_1,p_2|q_2,p_1\rangle=\frac{1}{2\pi\hbar}\exp[-\frac{i}{\hbar}(q_2
p_2 - \frac{\theta}{\hbar} p_1 p_2 -q_1 p_1)].\label{2.7} \ee

Making use of (\ref{2.7}) and the Baker-Campbell-Hausdorff (BCH)
theorem, it is fairly direct to show that \be
\frac{1}{(2\pi\hbar)^2}\Tr\{\exp[\frac{i}{\hbar}(({\bf y}-{\bf
y}^{\prime})\cdot {\hat{\bf R}} +({\bf x}-{\bf
x}^{\prime})\cdot{\hat {\bf P}})]\}=\delta({\bf x}-{\bf x}^{\prime})
\delta({\bf y}-{\bf y}^{\prime}),\label{2.8} \ee
where ${\bf x}= (x_1, x_2)\;\;\; {\bf y}= (y_1, y_2)$.\\
Thus for our extended Heisenberg algebra also the
$\{(2\pi\hbar)^{-1} \exp[\frac{i}{\hbar}({\bf y}\cdot {\bf\hat  R}
+{\bf x}\cdot{\bf\hat  P})]\}$ form a complete set of orthonormal
operators. and any Schr\"{o}dinger operator (which may depend
explicitly on time) $A({\bf \hat P}, {\bf \hat R}, t)$ can be
written as \be A({\bf \hat P}, {\bf \hat R}, t)= \int \int d{\bf x}
\ d{\bf y} \alpha({\bf x}, {\bf y}, t) \exp[\frac{i}{\hbar}({\bf
x}\cdot {\bf \hat P} + {\bf y}\cdot {\bf \hat R})],\label{op} \ee
where, by (\ref{2.8}), the $c$-function $\alpha({\bf x}, {\bf y},
t)$ is determined by \be \alpha({\bf x}, {\bf y},
t)=(2\pi\hbar)^{-2} \Tr \{ A({\bf \hat P}, {\bf \hat R}, t)
\exp[-\frac{i}{\hbar}({\bf x}\cdot {\bf \hat P}
 + {\bf y}\cdot {\bf \hat R})] \}.\label{alpha}
\ee The Weyl function corresponding to the quantum operator $A({\bf
\hat P}, {\bf \hat R}, t)$ is then given by \be
\begin{split}
A_{W} ({\bf p},{\bf q}, t)= \int \int d{\bf x} \ d{\bf y}
\;\alpha({\bf x},
{\bf y}, t) \exp[\frac{i}{\hbar}({\bf x}\cdot {\bf p} + {\bf y}\cdot {\bf q})]=\hspace{1in}\\
\int\int dx_1 dy_2 e^{\frac{i}{\hbar}(x_1 p_1 +y_2 q_2)}\langle q_1
-\frac{x_1}{2}-\frac{\theta y_2}{2\hbar}, p_2 +\frac{y_2}{2}|{\hat
A}|q_1 +\frac{x_1}{2}+\frac{\theta y_2}{2\hbar}, p_2
-\frac{y_2}{2}\rangle. \label{weyl}
\end{split}
\ee To derive the expectation value of a product of two
Schr\"{o}dinger operators, one writes the expectation value of the
product in terms of the von Neumann density matrix ${\boldsymbol
\rho}$  as \be \langle{\hat A}_1 {\hat A}_2 \rangle=\Tr[{\boldsymbol
\rho}{\hat A}_1 {\hat A}_2],\label{2.9} \ee and evaluates the trace
in the above chosen basis. After a rather lengthy but fairly
straightforward calculation the result obtained is \be
\begin{split}
\langle{\hat A}_1 {\hat A}_2 \rangle= \int\dots\int dp_1 dp_2 dq_1
dq_2 \; \frac{1}{(2\pi\hbar)^2}\int
d\xi d\eta e^{-\frac{i}{\hbar}(\eta q_2 -\xi p_1)}\\
\langle q_1 -\frac{\xi}{2}, p_2 -\frac{\eta}{2}|{\boldsymbol
\rho}|q_1 +\frac{\xi}{2}, p_2 +\frac{\eta}{2}\rangle
e^{\frac{1}{\hbar}\theta p_1 \partial_{q_2}}((A_1)_{W} \star_{\hbar}
(A_2)_W),\label{2.10}
\end{split}
\ee where \be \star_{\hbar}:=\exp [ \frac{i\hbar}{2} \Lambda ] :=
\exp \left[ \frac{i\hbar}{2} ( {\overleftarrow{\nabla_{{\bf
q}}}}\cdot {\overrightarrow{\nabla_{{\bf p}}}} -
{\overleftarrow{\nabla_{{\bf p}}}}\cdot
{\overrightarrow{\nabla_{{\bf q}}}} ) \right],\label{2.11} \ee is
the Gronewold-Moyal star-product bidifferential of the usual WWGM
quantum mechanics formalism. If we now let \be
\rho_{(Wigner)}:=\frac{1}{(2\pi\hbar)^2}\int
d\xi d\eta e^{-\frac{i}{\hbar}(\eta q_2 -\xi p_1)}\\
\langle q_1 -\frac{\xi}{2}, p_2 -\frac{\eta}{2}|{\boldsymbol
\rho}|q_1 +\frac{\xi}{2}, p_2 + \frac{\eta}{2}\rangle\label{2.12}
\ee denote the standard Wigner quasi-probability distribution in our
chosen basis, then (\ref{2.10}) reads as \be \langle{\hat A}_1 {\hat
A}_2 \rangle= \int\int d{\bf p} d{\bf q}\;\rho_{(Wigner)}
e^{\frac{1}{\hbar}\theta p_1 \partial_{q_2}}((A_1)_W \star_{\hbar}
(A_2)_W).\label{2.13} \ee Note that we could equally well have
integrated the above equation by parts to get \be \langle{\hat A}_1
{\hat A}_2 \rangle= \int \int d{\bf p} d{\bf q}\;\rho_{W} ((A_1)_W
\star_{\hbar} (A_2)_W).\label{2.14} \ee where the Weyl function
$\rho_{W}$ corresponding to ${\boldsymbol \rho}$ is related to
$\rho_{(Wigner)}$ by \be \rho_{W}= e^{-\frac{1}{\hbar}\theta p_1
\partial_{q_2}}(\rho_{(Wigner)}), \label{2.15} \ee in
contradistinction to what happens in the usual quantum mechanics
where they are the same. So in the calculation of the expectation
value of the product of two Schr\"{o}dinger operators, the
quantities that enter in the quantum mechanics based on the extended
Heisenberg algebra are either $((A_1)_W \star_{\hbar} (A_2)_W)$,
when averaging with $\rho_{W}$, or
 $e^{\frac{1}{\hbar}\theta p_1 \partial_{q_2}}((A_1)_W \star_{\hbar} (A_2)_W)$
when averaging with the usual Wigner function. However, also
contrary to what happens in ordinary quantum mechanics, these
quantities are not equal to the Weyl
equivalent $( {\hat A}_1 {\hat A}_2)_{W}$ of the product ${\hat A}_1 {\hat A}_2$.\\
To evaluate $( {\hat A}_1 {\hat A}_2)_{W}$ we use (\ref{alpha}) and
(\ref{weyl}), and following steps entirely analogous to the ones
treated in more detail in the following section when considering
Heisenberg operators, it can be shown that \be ( {\hat A}_1 {\hat
A}_2)_{W} = ({\hat A}_1)_W \bigstar({\hat A}_2)_W ,\label{weylprod}
\ee
where $\bigstar$ is defined by the composition of operator bi-differentials:
\be \bigstar:=\star_{\theta}\circ \star_{\hbar},\label{superstar}
\ee with $\star_{\hbar}$ as defined in (\ref{2.11}) and \be
\star_{\theta} := e^{\frac{i\theta}{2}({\overleftarrow
\partial}_{q_1}{\overrightarrow \partial}_{q_2} -{\overleftarrow
\partial}_{q_2}{\overrightarrow \partial}_{q_1})}.\label{startheta}
\ee

Furthermore and similarly to what occurs in ordinary quantum
mechanics, there is a stronger star-value equation related to
(\ref{2.14}). There are again however important differences. Thus,
given a Hamiltonian operator ${\hat H}$ and a pure energy state
satisfying the eigenvalue equation ${\hat H}
|\psi\rangle=E|\psi\rangle$, it can be shown that the star-value
equation for the quantum mechanics with our extended Heisenberg
algebra is \be {\bar H}_W \;\star_{\hbar} \;\rho_{(Wigner)} =
E\;\rho_{(Wigner)},\label{2.16} \ee where \be {\bar H}_{W} ({\bf p},
{\bf q}) = e^{\frac{1}{\hbar}\theta p_1 \partial_{q_2}} H_{W}({\bf
p}, {\bf q}).\label{barH} \ee Because of space limitations we omit
here the details of the proof of this theorem. These, together with
other more detailed aspects of our previous discussion as well
examples where specific implications of the quantum mechanics here
summarized are displayed and compared with other approaches, will be
dealt with in a
forthcoming paper to appear elsewhere.\\

\section{Weyl Equivalent of Heisenberg Operators}
Let \be \Omega^H :=\Omega({\bf \hat P}, {\bf \hat R},
t):=e^{\frac{it}{\hbar} {\hat H}} \Omega({\bf \hat P}, {\bf \hat R},
0) e^{-\frac{it}{\hbar} {\hat H}},\label{heisen} \ee
be the Heisenberg operator corresponding to the Schr\"{o}dinger operator $\Omega({\bf \hat P}, {\bf \hat R}, 0)$.\\
As for Schr\"{o}dinger operators the $c$-function
$\alpha_{\Omega}({\bf x}, {\bf y},t)$, associated with the Weyl
function $(\Omega^H)_W$ defined as in (\ref{weyl}), is given by (see
(\ref{alpha})) \be \alpha_{\Omega}({\bf x}, {\bf y},t)=
(2\pi\hbar)^{-2} \Tr\{e^{\frac{it}{\hbar} {\hat H}} \Omega({\bf \hat
P}, {\bf \hat R}, 0) e^{-\frac{it}{\hbar} {\hat H}}
e^{-\frac{i}{\hbar}({\bf x}\cdot{\bf \hat P}+{\bf y}\cdot{\bf \hat
R})}\}.\label{3.1} \ee Differentiating (\ref{weyl}) with respect to
$t$ and taking the Fourier transform gives immediately \be
\begin{split}
\frac{\partial \alpha_{\Omega}}{\partial t}=
\frac{i(2\pi\hbar)^{-2}}{\hbar}\int dq_1 dp_2 \langle
q_1-\frac{x_1}{2}-\frac{y_2 \theta}{2\hbar},
p_2 +\frac{y_2}{2}|[H,\Omega^H]| q_1+\frac{x_1}{2}+\frac{y_2 \theta}{2\hbar},p_2 -\frac{y_2}{2}\rangle\\
\times \exp[-\frac{i}{\hbar}(y_1 q_1+x_2
p_2)].\hspace{3in}\label{3.2}
\end{split}
\ee Consider now the quantity
$$\int dq_1 dp_2 \exp[-\frac{i}{\hbar}(y_1 q_1+x_2 p_2)]
\langle q_1-\frac{x_1}{2}-\frac{y_2 \theta}{2\hbar}, p_2
+\frac{y_2}{2}|H\Omega^{H}| q_1+\frac{x_1}{2}+\frac{y_2
\theta}{2\hbar},p_2 -\frac{y_2}{2}\rangle$$ which, after making use
of (\ref{op}), (\ref{2.7}), the BCH theorem and performing several
fairly direct integrations, yields \be
\begin{split}
(2\pi\hbar)^{-2}\int dq_1 dp_2 \exp[-\frac{i}{\hbar}(y_1 q_1+x_2
p_2)]&\\
 \langle q_1-\frac{x_1}{2}-\frac{y_2 \theta}{2\hbar}, p_2
+\frac{y_2}{2}|H\Omega^{H}| q_1+\frac{x_1}{2}+\frac{y_2
\theta}{2\hbar},p_2 -&\frac{y_2}{2}\rangle=\\
\int d{\bf x}^{\prime} d{\bf y}^{\prime} \alpha_{H}({\bf x}^{\prime}
{\bf y}^{\prime})
\alpha_{\Omega}({\bf x} -{\bf x}^{\prime}, {\bf y} -{\bf y}^{\prime},t)\\
\times\exp[\frac{i}{2\hbar} (-\frac{y_1^{\prime} y_2 \theta}{\hbar}
+ \frac{y_1 y^{\prime}_2 \theta}{\hbar}+
 x_2^{\prime} y_2 -x_2 y_2^{\prime}+ y_1 x_1^{\prime}-y_1^{\prime} x_1)].\label{3.3}
\end{split}
\ee Rewriting (\ref{3.3}) in terms of $H_W$ and $(\Omega^{H})_W$, by
making use of the Fourier inverse of the first equality in
(\ref{weyl}), and substituting the result into (\ref{3.2}) it
readily follows that \be
\begin{split}
\frac{\partial \alpha_{\Omega}}{\partial t}=
\frac{i(2\pi\hbar)^{-8}}{\hbar}\int\dots\int d{\bf p}^{\prime}d{\bf
q}^{\prime}d{\bf p}^{\prime\prime}d{\bf q}^{\prime\prime} d{\bf
x}^{\prime}d{\bf y}^{\prime}e^{-\frac{i}{\hbar}({\bf
x}^{\prime}\cdot{\bf p}^{\prime}
+ {\bf y}^{\prime}\cdot{\bf q}^{\prime})}\hspace{2in}\\
\times [H_{W}({\bf p}^{\prime},{\bf q}^{\prime})\Omega^{H}_{W}({\bf
p}^{\prime\prime}, {\bf q}^{\prime\prime},t) - \Omega^{H}_{W}({\bf
p}^{\prime},{\bf q}^{\prime})H_{W}({\bf p}^{\prime\prime},
{\bf q}^{\prime\prime},t)]\hspace{2in}\\
\times \exp[-\frac{i}{\hbar}(({\bf x}-{\bf x}^{\prime})\cdot{\bf
p}^{\prime\prime}
+ ({\bf y}-{\bf y}^{\prime})\cdot{\bf q}^{\prime\prime})]\hspace{2.9in}\\
\times\exp[\frac{i}{2\hbar} (-\frac{y_1^{\prime} y_2 \theta}{\hbar}
+ \frac{y_1 y^{\prime}_2 \theta}{\hbar}+ x_2^{\prime} y_2 -x_2
y_2^{\prime}+
 y_1 x_1^{\prime}-y_1^{\prime} x_1)]\hspace{1.8in}.\label{3.4}
\end{split}
\ee

Finally, double Fourier transforming both sides of (\ref{3.4}),
rearranging terms and performing the integrals, yields \be
\frac{\partial \Omega^{H}_W}{\partial t}=\frac{i}{\hbar}[H_{W}({\bf
p},{\bf q})
 \bigstar \Omega^{H}_{W}({\bf p},{\bf q})
-\Omega^{H}_{W}({\bf p},{\bf q})\bigstar H_{W}({\bf p},{\bf
q})].\label{3.5} \ee Note that by interchanging the ordering of the
Weyl functions in the second term inside the square brackets in
(\ref{3.5}), we alternatively have \be \frac{\partial
\Omega^{H}_W}{\partial t}=\frac{i}{\hbar}\;H_{W}[e^{\frac{i}{2}
(\hbar\Lambda +\theta \Lambda^{\prime})}-e^{-\frac{i}{2}
(\hbar\Lambda +\theta \Lambda^{\prime})}]\;\Omega^{H}_{W} =
-\frac{2}{\hbar}\;H_{W}\; \sin[\frac{1}{2}(\hbar\Lambda +\theta
\Lambda^{\prime})]\; \Omega^{H}_{W},\label{3.5b} \ee where \be
\Lambda:={\overleftarrow \nabla}_{\bf q}\cdot{\overrightarrow
\nabla}_{\bf p}- {\overleftarrow \nabla}_{\bf
p}\cdot{\overrightarrow \nabla}_{\bf q}, \ \ \ \
\Lambda^{\prime}:={\overleftarrow
\partial}_{q_1}\cdot{\overrightarrow \partial}_{q_2}-
{\overleftarrow \partial}_{q_2}\cdot{\overrightarrow
\partial}_{q_1}.\label{3.6} \ee Equation (\ref{3.5b}) can be formally
integrated to give \be \Omega^{H}_{W}({\bf p},{\bf
q},t)=\exp\{-\frac{2t}{\hbar}\;H_{W} \sin[\frac{1}{2}(\hbar\Lambda
+\theta \Lambda^{\prime})]\}\Omega_{W}({\bf p},{\bf q},0).
\label{3.7} \ee

Note that (\ref{3.5}) is in agreement with the derivation in
\cite{hu} for the time evolution of the Wigner function, although
the calculation there is somewhat circular from our point of view as
it assumes the $\star_{\theta}$-product
to be valid {\it ab initio}.\\
\section{Noncommutative Field Theory from extended Heisenberg algebra Quantum Mechanics}
Up to this point in the WWGM formalism the ${\bf
q}$'s and  ${\bf p}$'s (the
continuum of eigenvalues of ${\bf \hat R}$ and ${\bf \hat P}$) are only variables of integration . In
order to be able to interpret them as canonical dynamical variables,
as it is the case for ordinary WWGM quantum mechanics,
let us consider the specific cases when the Heisenberg operator
$\Omega^H$ in Section 3 is ${\bf \hat P}(t)$ or ${\bf \hat R}(t)$.
Making use of (\ref{weyl}) and (\ref{3.7}), and recalling that ${\bf
P}_W({\bf p},{\bf q},0)={\bf p}$ and ${\bf R}_W({\bf p},{\bf
q},0)={\bf q}$, we get for this particular cases, and a mechanical
Hamiltonian of the form ${\hat H}=\frac{{\bf \hat P}^2}{2m} + V({
\bf\hat R})$, \ba
\frac{d \bf P^{H}_{W}}{dt}|_{t=0} &=&-\frac{1}{\hbar}H(\hbar\Lambda +\theta\Lambda^{\prime}){\bf p}=-\nabla_{\bf q} V,\nonumber\\
\frac{d (R_1^{H})_{W}}{dt}|_{t=0} &=&=-\frac{1}{\hbar}H(\hbar\Lambda
+\theta\Lambda^{\prime}){ q_1}= \frac{p_1}{2m}
+\frac{\theta}{\hbar}\partial_{q_2}V,\nonumber\\
\frac{d (R_2^{H})_{W}}{dt}|_{t=0}&=& =
-\frac{1}{\hbar}H(\hbar\Lambda +\theta\Lambda^{\prime}){
q_2}=\frac{p_2}{2m} -\frac{\theta}{\hbar}\partial_{q_1}V. \label{PB}
\ea
Introducing now the following fundamental Poisson brackets as part of
the algebra structure of the ${\bf
q}$'s and  ${\bf p}$'s: \be \{ p_i,p_j\}=0, \ \ \ \{
q_i,q_j\}=\frac{\theta_{ij}}{\hbar}, \ \ \ \{
q_i,p_j\}=\delta_{ij},\label{4.2} \ee  we
have that (\ref{PB}) read \be \frac{d ( P_i^{H})_{W}}{dt}|_{t=0}
=\{p_i,H\}={\dot p}_i,\ \ \ \frac{d (R_i^{H})_{W}}{dt}|_{t=0}
=\{q_i,H\}={\dot q}_i,\label{4.3} \ee and therefore with this additional Poisson structure
the ${\bf q}$'s and
${\bf p}$'s satisfy the
Hamilton equations and can be considered formally as
canonical dynamical variables in the theory.

A representation for the above Poisson brackets can be constructed
by defining the twisted product \be q_i \star_{\theta} q_j := q_i
e^{\frac{i}{2}\sum_{lm}{\overleftarrow\partial}_{ q_l} \theta_{lm}
{\overrightarrow\partial}_{q_m}} q_j,\label{4.4} \ee where we have
generalized our arguments to ${\mathbb R}^d \:\:(\text
{with}\;\;d\ge 2) $, and letting \be \{ q_i,q_j\}:=
-\frac{i}{\hbar}[q_i,q_j]_{\star_{\theta}}: =-\frac{i}{\hbar}[q_i
\star_{\theta} q_j -q_j \star_{\theta} q_i].\label{4.5} \ee We can
consequently argue that the noncommutativiy of the extended
Heisenberg algebra in Quantum Mechanics manifests itself as a
twisting in the product of the algebra of the corresponding
classical canonical dynamical variables which, in accordance with
\cite{chaichian}, may be interpreted in turn as an Abelian Drinfeld
twisting of the coproduct in the Hopf algebra $\mathcal H$ of the
universal envelope $\mathcal U(\mathcal G)$ of the Galileo symmetry
algebra. If we now view the  module algebra $\mathcal A_{\theta}$
(the so called Groenewold-Moyal plane), described in the
Introduction, as a certain completion of the algebra generated by
the $q_i$ and describe fields as elements of $\mathcal A_{\theta}$,
then fields will clearly
inherit the $\star_{\theta}$-product.\\

As a final parenthetical remark, note from Sec 2 that in all the
expressions based on the WWGM formalism containing the $\theta$, it
always appears in the form of of the quotient
$\frac{\theta}{\hbar}$. If we claim that the noncommutativity
(\ref{noncomm1}) in the extended Heisenberg algebra is originated
from quantum gravity, then it is reasonable to assume (as already
mentioned in the Introduction) that $\theta\sim l_p^2
=\frac{k\hbar}{c^3}$, where $l_p$ is the Planck length and $k$ is
the gravitational coupling constant. Thus
$\frac{\theta^{ij}}{\hbar}\sim \frac{k}{c^3}$. This shows then that
corrections, due to this noncommutativity, to calculations such as
energy spectra and equations of motion such as (\ref{PB}),  are
indifferent to the value of $\hbar$, and that even in the limit
$\hbar\rightarrow 0$ there is what may appear  as a remanent of
quantum gravity.

\section*{Acknowledgments}
The authors acknowledge partial support from CONACyT projects
UA7899-F (M.R.) and 47211-F(J.D.V.) and  DGAPA-UNAM grant IN104503
(J.D.V.).

\end{document}